\documentclass[epj]{svjour}

\usepackage{graphics}

\newcommand{\be}{ \begin{equation}}
\newcommand{\ee}{\end{equation}} 

\begin{document}
\title{Gauge-invariant gravitational wave modes  in pre-big bang 
cosmology}
\author{Valerio Faraoni\inst{} 
}                     


\institute{Physics Department, Bishop's University, 
2600 College Street, Sherbrooke, 
Qu\'ebec, Canada J1M~1Z7\\
email vfaraoni@ubishops.ca}

\date{Received: date / Revised version: date}

\abstract{The $t<0$ branch of pre-big bang cosmological 
scenarios is  subject to a gravitational wave instability. The  
unstable 
behaviour of tensor perturbations is derived in a very 
simple way in Hwang's covariant and  gauge-invariant formalism 
developed for extended  theories of gravity. A simple 
interpretation of this instability as the effect of an 
``antifriction'' is given , and it is argued that a universe 
must eventually enter the expanding phase.
} 

\maketitle

\section{Introduction}
\label{intro}

The application of string theory to cosmology has produced, among 
other things, pre-big bang cosmology (see 
\cite{Gasperinibook,Gasperini} 
for a recent exhaustive review). Low-energy string cosmology can 
be described by the action of the graviton-dilaton sector 
\be
S=\frac{1}{2\lambda_S^{d-1}} \int d^{d+1}x \sqrt{-g} \, 
\mbox{e}^{-\Phi} \left( 
R+g^{\mu\nu}\nabla_{\mu}\Phi\nabla_{\nu}\Phi \right)
\ee
plus matter, where $\lambda_S$ is the natural unit of length, 
$g_{\mu\nu}$ is the metric tensor with determinant $g$, and $R$ 
is the Ricci curvature. Two features are characteristic of string 
theories 
and unavoidable: the first is the presence of the dilaton field 
$\Phi $  together with the metric tensor 
$g_{\mu\nu}$, and the second is the non-minimal coupling  of 
$\Phi$ to the Ricci scalar $R$. Following \cite{Gasperinibook}, 
we neglect for simplicity the matter sector and focus on the 
cosmological effects of the dilaton. Assuming the extra spatial 
dimensions to be compactified, the action can be reduced to the 
Brans-Dicke form
\be
S_{BD}=\int d^4x \sqrt{-g} \left( \frac{ \varphi R}{2} -\frac{ 
\omega_0}{\varphi} \, 
g^{\mu\nu} \nabla_{\mu}\varphi\nabla_{\nu}\varphi 
\right) \label{BransDickeaction}
\ee
provided that $\omega_0=-1$ (it is well know that the low-energy 
limit of the bosonic string, the prototype of string theories, 
leads to Brans-Dicke gravity with Brans-Dicke parameter 
$\omega_0=-1$ \cite{bosonicstring}). 

In pre-big bang cosmology it is common to use the shifted dilaton 
$\bar{\Phi} \equiv \Phi-\ln \left( \Pi_ia_i \right)$, where the 
$a_i$ are the scale factors of the $d$-dimensional Bianchi metric 
$ds^2=-dt^2+\sum_i a_i^2 (dx^i)^2$. Here we consider spatially 
flat Friedmann-Lemaitre-Robertson-Walker (FLRW) metrics, which 
are the most recurrent in pre-big bang cosmology, and 
$d=3$, the extra spatial dimensions being compactified, therefore 
$\bar{\Phi}=\Phi-3\ln a$, where $a$ is the scale factor of the 
three-dimensional spatial sections and 
\be
 \varphi = \mbox{e}^{-\Phi}=\frac{ \mbox{e}^{-\bar{\Phi} }}{a^3} 
\,.
\ee
The following solutions derived using scale factor duality 
transformations \cite{duality} are representative of pre-big bang 
cosmology \cite{Gasperinibook,Gasperini}:
\begin{eqnarray}
a_{(1)}(t) & = & t^{1/\sqrt{3}} \,, \;\;\;\;  
\bar{\Phi}_{(1)}(t) = -\ln t \;\;\;\; (t> 0),\\
&& \nonumber\\
a_{(2)}(t) & = & (-t)^{1/\sqrt{3}} \,, \;\;\;\;\;  
\bar{\Phi}_{(2)}(t) = -\ln (- t) \;\;\;\;\; (t<0), \nonumber\\
&& \\
a_{(3)}(t) & = & t^{-1/\sqrt{3}} \,, \;\;\;\;\;  
\bar{\Phi}_{(3)}(t) = -\ln t \;\;\;\;\; (t>0),\\
&& \nonumber\\
a_{(4)}(t) & = & (-t)^{-1/\sqrt{3}} \,, \;\;\;\;\;  
\bar{\Phi}_{(4)}(t) = -\ln (-t) \;\;\;\;\; (t<0) . \nonumber\\
&&
\end{eqnarray}
There is a singularity at $t=0$, which is hoped to be 
somehow eliminated due to the existence of a fundamental minimum 
length, the size  of the string, so that the two branches of pre- 
and post-big bang  can be joined smoothly through a bounce or a 
``graceful exit'' \cite{Gasperini,Gasperinibook}.

The stability of these solutions is an issue.  Gravitational 
instabilities are well known to  plague pre-big bang cosmology 
\cite{Gasperinibook} and here we  want to study tensor 
perturbations and their stability.   Inhomogeneous  
perturbations in cosmology are subject to the 
notorious  gauge-dependence problems and need to be analyzed in 
a  gauge-invariant way. In the effective Brans-Dicke 
theory~(\ref{BransDickeaction}), inhomogeneous perturbations are 
best analyzed using the 
covariant and gauge-invariant formalism of Bardeen, Ellis and 
Bruni \cite{Bardeen,EllisBruni} in the version developed 
by Hwang  and Noh \cite{Hwang} for alternative gravity 
described by the 
action
\be
S=\int d^4x \, \sqrt{-g} \, \left[ \frac{f \left(\phi, R 
\right)}{2} 
-\frac{ \omega(\phi)}{2}  \, 
g^{\mu\nu}\nabla_{\mu}\phi\nabla_{\nu}\phi -V(\phi) \right] \,.
\ee
This formalism is particularly convenient to analyze 
perturbations in the pre-big bang 
scenarios described by~(\ref{BransDickeaction}).  In the next 
section, the basic variables and equations 
of this formalism are rewieved and applied to pre-big bang 
cosmology.

\section{The gauge-invariant formalism}

The metric components are  written as 
\begin{eqnarray} 
g_{00} & = & -a^2 \left( 1+2AY \right) \,, \ \\
&& \nonumber \\
g_{0i} & = & -a^2 \, B \, Y_i  \,,  \\
&& \nonumber \\
g_{ij} & =& a^2 \left[ h_{ij}\left(   1+2H_L \right) +2H_T \, 
Y_{ij}  \right] \,,
\end{eqnarray}
where $h_{ij} $ is the three-dimensional metric of the FLRW 
background with associated covariant derivative $ 
\bar{\nabla_i} $. The scalar harmonics 
$Y$ are the eigenfunctions of the eigenvalue problem
\be 
\bar{\nabla_i} \bar{\nabla^i} \, Y =-k^2 \, Y \,,
\ee
with eigenvalue $k$, while the vector and tensor 
harmonics $Y_i$ and $Y_{ij}$ satisfy
\be 
Y_i= -\frac{1}{k} \, \bar{\nabla_i} Y \,, \;\;\;\;\;\;\;\;
Y_{ij}= \frac{1}{k^2} \, \bar{\nabla_i} \bar{\nabla_j} Y 
+\frac{1}{3} \, Y \, h_{ij} \,.
\ee
The gauge-invariant variables used in Hwang's formalism are the 
Bardeen \cite{Bardeen} potentials 
\begin{eqnarray} 
\Phi_H & = & H_L +\frac{H_T}{3} +\frac{ \dot{a} }{k} \left( 
B-\frac{a}{k} \, \dot{H}_T \right) \,, \label{phiH}\\
&&\nonumber\\
\Phi_A & = &  A  +\frac{ \dot{a} }{k} \left( B-\frac{a}{k} \, 
\dot{H}_T \right)
+\frac{a}{k} \left[ \dot{B} -\frac{1}{k} \left( a \dot{H}_T 
\right)\dot{}  \right] \,,  \nonumber\\
&& \label{phiA}
\end{eqnarray}
and the  Ellis-Bruni \cite{EllisBruni} variable 
\be \label{deltaphi}
\Delta \phi = \delta \phi  +\frac{a}{k} \, \dot{\phi}  \left( 
B-\frac{a}{k} \, \dot{H}_T 
\right) \,,
\ee
where $a$ is the scale factor of the background FLRW line 
element 
\be
ds^2=-dt^2+a^2(t)\left( dx^2+dy^2+dz^2 \right)
\ee
and an  overdot denotes differentiation with respect to the 
comoving  time  $t$. The first order equations for the 
gauge-invariant perturbations 
are \cite{Hwang}
\begin{eqnarray}  
 \Delta \ddot{\phi} & + &  \left( 3H + \frac{\dot{\phi}}{ 
\omega} \, 
\frac{d\omega}{d\phi} \right) \Delta 
\dot{\phi} \nonumber\\
&&\nonumber\\
& +& \left[ \frac{k^2}{a^2}
+ \frac{\dot{\phi}^2}{2} \frac{d}{d\phi} \left( 
\frac{1}{\omega} \frac{d\omega}{d\phi} \right) -\, 
\frac{d}{d\phi} \left( \frac{1}{2\omega} \frac{\partial 
f}{\partial \phi} -\frac{1}{\omega} 
\frac{dV}{d\phi} \right) \right] \Delta \phi  \nonumber \\
&& \nonumber \\
& = &
 \dot{\phi}  \left(  \dot{\Phi}_A - 3\dot{\Phi}_H \right) 
+ \frac{\Phi_A}{\omega}  \left( \frac{\partial f}{\partial 
\phi} -2 \, \frac{dV}{d\phi} \right) \nonumber\\
&&\nonumber\\ 
& + & \frac{1}{2\omega} \, \frac{\partial^2 f}{\partial \phi 
\partial  
R}  \, \Delta R  \, ,\nonumber\\
&&
\end{eqnarray}
\begin{eqnarray}  
 \Delta \ddot{F} & + & 3H \Delta \dot{F} +\left( \frac{k^2}{a^2} 
-  \frac{R}{3} \right) \Delta F  +\frac{F}{3} \, \Delta R 
+\frac{2}{3} \, \omega \dot{\phi} \Delta \dot{\phi} \nonumber\\
&&\nonumber\\
& + & \frac{1}{3} \left( \dot{\phi}^2 \frac{d\omega}{d\phi} + 
2\frac{\partial f}{\partial \phi} 
-4 \, \frac{dV}{d\phi} \right) \Delta \phi \nonumber \\
&& \nonumber \\
&& = \dot{F}  \left(  \dot{\Phi}_A - 3\dot{\Phi}_H \right) 
+ \frac{2}{3}  \left( FR -2f +4V  \right)  \Phi_A \, ,
\end{eqnarray}
\be \label{HT}
\ddot{H}_T +\left( 3H+ \frac{\dot{F}}{F} \right) \dot{H}_T 
+\frac{k^2}{a^2} \, H_T=0 \,,
\ee
\be  
- \dot{\Phi}_H +\left( H +  \frac{\dot{F}}{2F} \right) \Phi_A = 
\frac{1}{2} \left(  \frac{  \Delta \dot{F} }{F} -H \frac{ \Delta F}{F} +  
\frac{ \omega}{F} \,  \dot{\phi} \, \Delta \phi \right)  \, ,
\ee
\begin{eqnarray} 
& \, & \left( \frac{k}{a} \right)^2 \Phi_H +\frac{1}{2}
\left( \frac{ \omega}{F } \dot{\phi}^2  
+ \frac{3}{2} \frac{\dot{F}^2}{F^2}  \right) \Phi_A \nonumber\\
&&\nonumber\\
& = & 
\frac{1}{2} \left\{ \frac{3}{2} \frac{ \dot{F} \Delta \dot{F} 
}{F^2} + \left( 3\dot{H} -  
\frac{k^2}{a^2} -\frac{3H}{2} \frac{ \dot{F}}{F}  \right)  
\frac{ \Delta F}{F}   +\frac{\omega}{F} \dot{\phi} \Delta 
\dot{\phi} \right.\nonumber\\
&&\nonumber\\
& + & \left. \frac{1}{2F} \left[  \dot{\phi}^2 
\frac{d\omega}{d\phi} - \frac{ \partial f}{\partial \phi} 
+2\frac{dV}{d\phi} + 6\omega \dot{\phi} \left( H +  \frac{  
\dot{F} }{2F} \right) \right] \Delta \phi  \right\} 
\,,\nonumber\\
&&
\end{eqnarray}
\be  
\Phi_A + \Phi_H =  - \frac{\Delta F }{F} \, ,
\ee
\begin{eqnarray}  
\ddot{\Phi}_H & + & H \dot{\Phi}_H + \left( H + \frac{ 
\dot{F}}{2F} \right) 
\left( 2\dot{\Phi}_H -\dot{\Phi}_A \right) \nonumber\\
&&\nonumber\\
& + & \frac{ 1 }{2F}  \left( f-2V -RF \right)  \Phi_A \nonumber 
\\ 
&& \nonumber \\
& = & - \frac{1}{2} \left[ 
\frac{  \Delta \ddot{F}}{F} + 2H \, \frac{\Delta \dot{F}}{F} 
+ \left( P-\rho \right) \frac{ \Delta F}{2F} + \frac{ \omega}{F} \, \dot{\phi} \, \Delta 
\dot{\phi} \right.\nonumber\\
&&\nonumber\\
&+ & \left. \frac{1}{2F} \left( \dot{\phi}^2 \, \frac{d\omega}{ 
d\phi}  +\frac{\partial 
f}{\partial \phi  }  -2 \, \frac{dV}{d\phi } \right) \Delta \phi  
\right]  \, , \nonumber\\
&&  
\end{eqnarray}
\begin{eqnarray}
\Delta R & = & 6 \left[ \ddot{\Phi}_H +4H\dot{\Phi}_H 
+\frac{2}{3} \frac{k^2}{a^2} \Phi_H -H\dot{\Phi}_A  
\right.\nonumber\\
&&\nonumber\\
& - & \left. \left( 2\dot{H}+4H^2 -\frac{k^2}{3a^2} \right) 
\Phi_A 
\right] \,, 
\end{eqnarray}
where $F \equiv \partial f/\partial R$ and $ F_c \equiv h_c^d \, 
\nabla_d F$ is the spatial projection 
of the gradient of $F$ and the effective energy density and 
pressure of the scalar are 
\begin{eqnarray}  
\rho &=& \frac{1}{F}  \left[  \frac{\omega \, \dot{\phi}^2}{2}  
+\frac{  \left( RF -f +2V \right) }{2} -3H \dot{F} +  \nabla^c 
F_c \right] \, , \\
&&\nonumber\\
P &= & \frac{1}{F}  \left[  \frac{\omega \, \dot{\phi}^2}{2} 
+\frac{  \left( f -RF-2V \right)}{2} +\ddot{F} +2H \dot{F} 
-\frac{2}{3} \, 
\nabla^c F_c \right] \, .\nonumber\\
&&
\end{eqnarray}
While the analytic solution of the perturbation equations in a  
background  other than de Sitter \cite{deSitter} is a daunting 
task,  eq.~(\ref{HT})  
for the tensor modes $H_T$ is particularly simple since 
it is decoupled from the equations for the other modes. In the 
zero-momentum limit $k\rightarrow 0 $ it reduces to 
\be \label{HT2}
\ddot{H}_T+\left( 3H+\frac{\dot{\varphi}}{\varphi} \right) 
\dot{H}_T=0 
\ee
using the fact that $F=\varphi$. Note also that, using the well 
known result of pre-big bang cosmology $\dot{ \bar{\Phi}}=\pm 
\sqrt{3}\, H$ (see, {\em e.g.},  p.~139 of \cite{Gasperinibook}), 
the 
coefficient 
of $\dot{H}_T$ in eq.~(\ref{HT}) is $\left( 
3H+\dot{\varphi}/\varphi \right)=-\dot{ \bar{\Phi}}$ 
and corresponds 
to a friction term if $\dot{ \bar{\Phi}} <0 $
and to  antifriction if $\dot{ \bar{\Phi}} >0 $. Positive 
friction will damp the tensor perturbations, while antifriction 
enhances them.  Eq.~(\ref{HT2}) yields 
\be
\dot{H}_T=\frac{C}{a^3\varphi} \,,
\ee
where $C$ is an integration constant. 

Let us consider the contracting pre-big bang solution 
$  a_{(2)}(t)=\left( -t \right)^{1/\sqrt{3}}$, $ \bar{ 
\Phi}_{(2)}  (t)=-\ln (-t)$ for  $ t < 0$: here $ \dot{ 
\bar{\Phi}}   =-1/t >0 $ and 
\be
\dot{H}_T=\frac{C}{a^3\varphi} =C\, \mbox{e}^{ 
\bar{\Phi}}=\frac{C}{-t} 
\ee
diverges as $t\rightarrow 0^{-}$ (for general non-zero values of 
the arbitrary integration constant $C$). Similarly, for 
the pole-like expanding pre-big bang  branch $ a_{(4)}(t)=\left( 
-t  \right)^{-1/\sqrt{3}} $, $\bar{\Phi}_{(4)}=-\ln ( -t) $ 
(with $t<0$), the tensor perturbations diverge as $t\rightarrow 
0^{-}$. For both of these solutions $\dot{\bar{\Phi}}=-1/t>0$ 
corresponds to antifriction.

Let us consider now the post-big bang ($t>0$) branch 
$a_{(1)}(t)=t^{1/\sqrt{3}}$, $ \bar{\Phi}_{(1)}=-\ln t $; here 
\be
\dot{H}_T=C\, \mbox{e}^{\bar{ \Phi}} =\frac{C}{t} 
\ee
tends to zero as $t\rightarrow +\infty$ irrespective of the 
value of $C$ and $H_T$ becomes constant, corresponding to neutral 
stability. Note that in this case $ \dot{\bar{\Phi}} =-1/t <0$ 
corresponds to (positive) friction. The same considerations apply 
to $a_{(3)}(t)=t^{-1/\sqrt{3}} $, $\bar{\Phi}_{(3)}=-\ln t $ 
(for $t>0$).

\section{Discussion}

The pre-big bang branch is unstable with respect to long 
wavelength tensor perturbations. This fact is well 
known, but it appears in a particularly simple way in 
Hwang's formalism and the discussion is not tied to a 
particular gauge, as is often the case in the 
literature. Moreover, in the previous discussion, the 
instability receives a particularly simple  interpretation as 
the effect of an  antifriction in the evolution 
equation for these gauge-invariant modes, which is inherently 
due to the fact that the universe contracts.  This instability 
could  be regarded as a shortcoming of pre-big bang scenarios. 
However, it is also possible that breaking the evolution of the 
pre-big bang branch by means of tensor perturbations (of 
classical or quantum origin) that grow and drive the universe 
away for the background evolution helps in solving the 
long-standing graceful exit problem of joining the pre- and the 
post-big bang  branches. This possibility deserves attention in 
the future. In any case, it seems that a  contracting phase 
cannot last, eventually becoming unphysical,  and that the 
universe is bound to evolve into  an expanding one.

Another aspect of the gravitational wave instability is that, since 
long wavelength modes can grow, if they are somehow stopped before 
causing the dynamics to deviate drastically from a  FLRW one (which 
can only be assessed in a more detailed scenario and ideally  
requires to be confirmed by a second order perturbation analysis), 
large-scale  perturbations in the cosmic microwave background would be 
induced. 
While the experimental error in the observations is necessarily large 
at these large scales, there is hope that a signature of pre-big bang 
physics is still left in the cosmic microwave background. Although 
stabilization seems a problem at present, the possibility of such a 
signature makes it worth to pursue this possibility. It 
is not excluded that, in this case,  the growing $H_T$ modes induce 
effects 
detectable already by the {\em PLANCK} satellite \cite{PLANCK}. More 
detailed pre-big bang scenarios stabilized by extra ingredients will 
be analyzed (also at smaller scales) in future work  with 
this possibility in mind.

\section*{Acknowledgments} We thank a referee for useful comments. 
This  work is supported by the Natural Sciences and  Engineering 
Research Council of Canada  (NSERC).

\end{document}